\documentclass[prl,superscriptaddress,twocolumn]{revtex4}

\usepackage[dvipdfmx]{graphicx}
\usepackage{amssymb}
\usepackage{natbib}
\newcommand{\ped}[1]{\ensuremath{_{\rm #1}}}
\newcommand{\apex}[1]{\ensuremath{^{\rm #1}}}
\usepackage{color}

\begin{document}
\title{Control of bulk superconductivity in a BCS superconductor by surface charge doping via electrochemical gating}
\author{E. Piatti}
\author{D. Daghero}
\author{G. A. Ummarino}
\author{F. Laviano}
\author{J. R. Nair}

\affiliation{Department of Applied Science and Technology, Politecnico di Torino,
Torino, Italy }

\author{R. Cristiano}

\affiliation{CNR-SPIN Institute of Superconductors, Innovative Materials and Devices,
UOS-Napoli, Napoli, Italy}

\author{A. Casaburi}
\affiliation{School of Engineering, University of Glasgow, Glasgow, UK}

\author{C. Portesi}
\author{A. Sola}
\affiliation{INRIM - Istituto Nazionale di Ricerca Metrologica, Torino, Italy}

\author{R. S. Gonnelli}
\email{renato.gonnelli@polito.it}

\affiliation{Department of Applied Science and Technology, Politecnico di Torino,
Torino, Italy }

\date{\today}
\begin{abstract}
The electrochemical gating technique is a powerful tool to tune the \textit{surface} electronic conduction properties of various materials by means of pure charge doping, but its efficiency is thought to be hampered in materials with a good electronic screening. We show that, if applied to a metallic superconductor (NbN thin films), this approach allows observing reversible enhancements or suppressions of the \emph{bulk} superconducting transition temperature, which vary with the thickness of the films. These results are interpreted in terms of proximity effect, and indicate that the effective screening length depends on the induced charge density, becoming much larger than that predicted by standard screening theory at very high electric fields.
\end{abstract}


\maketitle

The field effect (i.e. the modulation of the conduction properties of a material by means of a static transverse electric field) is widely used in semiconducting electronic devices, namely FETs. Recently, unprecedented intensities of the electric field -- and thus densities of induced charge -- have been reached by exploiting the formation of an electric double layer (EDL) at the interface between an electrolyte and the solid, when a voltage is applied between the latter and a gate electrode immersed in the electrolyte. The EDL acts as a nanoscale capacitor with a nanometric spacing between the ``plates'', so that the electric field can be orders of magnitude higher than in standard field-effect (FE) devices.
In these extreme conditions, new phases (including superconductivity) have been discovered in various materials, mostly semiconducting or insulating in their native state \cite{Jo15,Ueno08,Ye12,Ye10,Ueno11}. Instead, high-carrier-density systems such as metals and standard BCS superconductors have so far received little attention, because the electronic screening strongly limits the FE. To the best of our knowledge, a few works on gold \cite{daghero12,nakayama12} and other noble metals \cite{tortello13} remain the only literature about EDL gating on normal metals. {\color{black}More exotic metallic systems, such as 2D materials from different classes \cite{shiogai15,lei16,xi16,yoshida16,li_nature16} and a variety of complex oxides \cite{bollinger11,leng11,leng12,maruyama15,jin16,walter16}, were explored more extensively. In particular, the microscopic mechanism behind the carrier density modulation in EDL-gated oxides remains a subject of investigation \cite{walter16,jeong13,li_acsnano13,schladt13}.}

The field effect on BCS superconductors was investigated in the Sixties via solid dielectric \cite{Glover60} and ferroelectric \cite{Stadler65} gating, and small (positive or negative) variations of the superconducting transition temperature ($T_c$) were observed on increasing/decreasing the charge carrier density. Similar results were recently obtained by EDL gating in Nb \cite{Choi14} thin films. In this case completely reversible $T_{c}$ shifts were observed, about three orders of magnitude larger than in \cite{Glover60,Stadler65}, though still smaller than 0.1 K. Despite the very effective electronic screening expected close to $T_c$ (and due to unpaired electrons) the suppression of $T_{c}$ was visible also in films as thick as 120 nm. This means that the superconducting properties of the \emph{bulk} were somehow changed by the applied gate voltage; otherwise, the surface layer with reduced $T_c$ would have been shunted by the underlying bulk giving no visible effect on the transition. A proper understanding of how this could happen is however still lacking \cite{Choi14}.

{\color{black}In this work we suggest a solution to this problem -- that first appeared in literature more than 50 years ago \cite{Glover60,Stadler65} --} by systematically studying the $T_c$ modulation of NbN thin films under EDL gating for different values of the film thickness $t$. We find that the $T_c$ shift {\color{black}(always smaller than 0.1 K, but its amplitude is not the main result of the paper)} does depend on $t$, thus proving that the whole bulk comes into play. We show that, if the proximity effect is taken into account in the strong-coupling limit of the standard BCS theory, this finding turns out to be compatible with a charge induction limited to the surface.  Interestingly, {\color{black}a fundamental quantity such as} the effective electrostatic screening length increases with the induced {\color{black}surface charge $\Delta n_{2D}$}, finally becoming much higher than the Thomas-Fermi screening length $\lambda_{TF}$ in the normal state {\color{black}when $\Delta n_{2D} > 2 \times 10^{15} \mathrm{cm^{-2}}$}.

\begin{figure}
\includegraphics[angle=0,width=0.75\columnwidth]{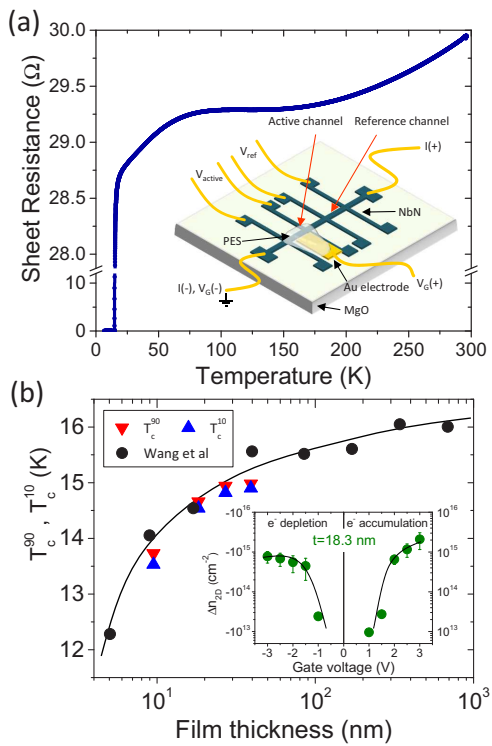}
\caption{(a) Sheet resistance as a function of temperature for the pristine 39.2 nm-thick device (prior to the PES deposition). The inset shows a schematic of the complete device. (b) Transition temperature as a function of thickness for our devices: both T\ped{c}\apex{90} (down triangles) and T\ped{c}\apex{10} (up triangles) are reported to show the variation in the transition width. Black dots are data taken from literature \cite{wang96}. Inset: typical $\Delta n_{2D}$ vs. $V_G$ curve determined by chronocoulometry.}\vspace{-5mm}
\label{figure:1}
\end{figure}

NbN thin films were grown on insulating MgO substrates by reactive magnetron sputtering. The device geometry was defined by photolithography and subsequent wet etching in a 1:1 HF:HNO\ped{3} solution. The inset to Fig. \ref{figure:1}a shows the scheme of the samples: the strip is 135 $\mu$m wide, with current pads on each end and four voltage contacts on each side, spaced by 946 $\mu$m from one another. This geometry allows measuring the voltage drop across different portions of the strip at the same time, and thus defining both an \emph{active} (gated) and a \emph{reference} (ungated) channel.

The thickness $t$ of the film was measured by atomic force microscopy (AFM). Fig. \ref{figure:1}a shows the sheet resistance $R_{\Box}$ vs temperature of the pristine film ($t=39.2 \pm 0.8$ nm). The non-monotonic behavior of $R_{\Box}(T)$ and the residual resistivity ratio $RRR = R(300 \mathrm{K})/R(16 \mathrm{K})= 1.05$ are characteristic of granular NbN films of fairly high quality \cite{Nigro98}. Subsequent steps of Ar-ion milling were used to progressively reduce the film thickness to $27.1 \pm 1.5$ nm,  $18.3 \pm 1.7$ nm and finally $9.5 \pm 1.8$ nm (see Supplemental Material for further details). On reducing $t$, the $T_c$ was progressively suppressed (in good agreement with the curve for NbN films reported in literature \cite{wang96}, see Fig. \ref{figure:1}b) and the transition width slightly increased. Both these effects are consistent with the fact that $t$ approaches the coherence length of the material \cite{wang96}.

To perform EDL gating measurements, we covered the active channel and the gate counterelectrode placed on its side (made of a thin Au flake: see inset to Fig. \ref{figure:1}a) with the liquid precursor of the cross-linked polymer electrolyte system (PES), which was later UV-cured.

As well known, Nb-based compounds always present a thin oxide layer at the surface (see \cite{semenov09} and references therein) that could be detrimental to EDL gating. However, for NbN this layer is less than 1 nm thick \cite{semenov09}, and does not significantly reduce the gate capacitance. Indeed, successful EDL gating through thin insulating layers was already reported \cite{gallagher16}; it actually turns out that the insulating layer minimizes the (unwanted) electrochemical reactions between sample and electrolyte.

To determine the surface electron density $\Delta n_{2D}$ induced at the surface by a gate voltage $V_G$, we used the well-established electrochemical technique called Double-Step Chronocoulometry \cite{Inzelt}. We applied a given $V_G$ \emph{at room temperature} (above the glass transition of the PES, which occurs below 230 K) as a step perturbation, and then removed it. As explained in Ref. \onlinecite{Piatti16}, an analysis of the gate current (as a function of time) allowed us to separate the contribution to the current due to diffusion of electroreactants from that due to the EDL build-up; from the latter, we can determine the charge stored in the EDL and thus $\Delta n_{2D}$. A typical $\Delta n_{2D}$ vs. $V_{G}$ curve is shown in the inset to Fig.\ref{figure:1}b. {\color{black}The reproducibility of the $\Delta n\ped{2D}$ estimation for multiple subsequent applications of the same $V_{G}$ is within $\sim$ 30$\%$ of the value, and is comparable with the uncertainty of the technique itself \cite{daghero12}.}

To measure the effect of a given $V_G$ on the transition temperature, we applied $V_G$ at room temperature and kept it constant while cooling the device down to 2.7 K in a pulse-tube cryocooler. Then, the voltage drops across the active and the reference channel, $V_{\mathrm{active}}$ and $V_{\mathrm{ref}}$ (see inset to Fig.\ref{figure:1}a) were measured simultaneously during the very slow, quasistatic heating up to room temperature in the presence of a source-drain dc current of a few $\mu$A.

The double-channel measurement allowed us to eliminate the possible small differences in critical temperature measured in different heating runs. By comparing the $R_{\Box}(T)$ curve of the active channel with that of the reference channel measured at the same time, we were able to detect shifts in $T_c$ as small as a few mK. For instance, the $T_c$ shift due to $V_G= +3$ V was evaluated as $\Delta T_c (3\, \mathrm{V})= [T_c^{active}-T_c^{ref}]_{V_G=3 \,V}-[T_c^{active}-T_c^{ref}]_{V_G=0\,V}$.

\begin{figure}
\includegraphics[angle=-90,width=0.75\columnwidth]{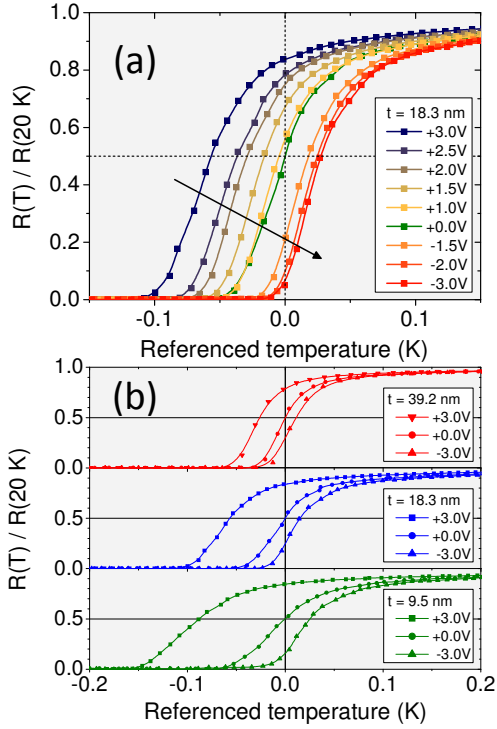}
\caption{(a) Normalized resistance $R(T)/R(20 \mathrm{K})$ of the active channel of a 18.3 nm thick device, as a function of referenced temperature $T^*$, i.e. $T^*=[T^{active}-T_c^{ref}]_{V_G}- [T_c^{active}-T_c^{ref}]_{0}$, at different gate voltages in the range [-3 V, +3 V]. (b) Effect of a gate voltage $V_G = \pm 3\mathrm{V}$ on the $R(T)/R(20\mathrm{K})$ vs. $T$ curve for three values of thickness: 39.2 nm, 18.3 nm and 9.5 nm.}\vspace{-5mm}
\label{figure:2}
\end{figure}

Fig. \ref{figure:2}a shows, as an example, the effect of a gate voltage ranging between +3 V and -3 V on the superconducting transition of the 18.3 nm thick film. The horizontal scale is the temperature normalized to the midpoint of the transition in the reference channel, i.e. $[T^{active}-T_c^{ref}]_{V_G}-[T_c^{active}-T_c^{ref}]_{0}$. As for all thicknesses, the  gate voltage reproducibly produces a \emph{rigid shift} of the superconducting transition to a lower (higher) temperature for positive (negative) $V_G$, respectively. The amplitude of the fully reversible shift (see Supplemental Material) is clearly correlated with the induced charge density.

\begin{figure}
\includegraphics[angle=-90,width=0.75\columnwidth]{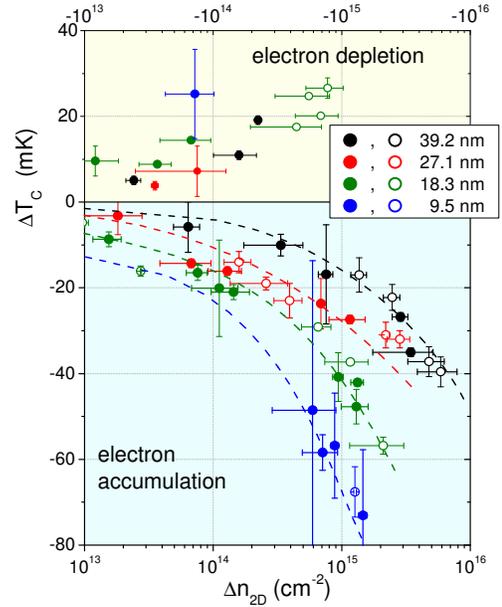}
\caption{$T_c$ shift, $\Delta T_c$, as a function of the induced surface electron density $\Delta n_{2D}$, for all the film thicknesses. Dashed lines are only guides to the eye.}\vspace{-5mm}
\label{figure:3}
\end{figure}

Figure \ref{figure:2}b shows that the amplitude of the $T_c$ shift produced by a given gate voltage (here $+ 3.0 \, \mathrm{V}$ and $- 3.0 \, \mathrm{V}$) is greatly enhanced when the thickness $t$ of the film is reduced. This fact (together with the detection of \emph{negative} shifts of $T_c$ for positive $V_G$) {\color{black}suggests} that the superconducting properties of the \emph{whole} bulk are affected by the surface charge induction. The values of $\Delta T_c$ as a function of $\Delta n_{2D}$ for the different thicknesses are shown in Fig. \ref{figure:3}.

Interestingly, the transition width depends on the film thickness but \emph{not} on the gate voltage, indicating that the charge induction does not create a $T_c$ gradient in the depth of the film. The question then is how the electric field can homogeneously perturb the superconducting properties in the whole thickness even in the presence of a strong electronic screening.

In general, describing the electrostatic screening in a superconductor is a complicated task in which various aspects should play a role. But in the proximity of $T_c$ the screening is dominated by unpaired electrons since the superfluid density is small. A two-fluid model in which the screening of Cooper pairs is calculated within the theory by Ovchinnikov \cite{Ovchinnikov77} gives that, 100 mK below $T_c$ (more than any shift we measured) the effective screening length is still 1.05 times the screening length of NbN in the normal state, that is of the order of 1 {\AA}.

{\color{black}The most likely mechanism able to turn the perturbation of the carrier density in a thin surface layer into a homogeneous perturbation of the bulk superconducting properties is the proximity effect at a normal metal/superconductor interface. In general, this is observed as the induction of a superconducting order parameter in the normal bank (close to the interface) accompanied by its suppression in the superconducting one \cite{DeGennesRMP1964}. However, when the thickness of both the surface layer and of the underlying bulk are smaller than the temperature-dependent coherence length $\xi(T)$ (Cooper limit \cite{CooperPRL1961}) the compound slab behaves as a homogeneous superconductor whose effective electron-phonon coupling constant responsible for the pairing is a weighted average of the coupling constants in the superconductor and in the normal metal \cite{CooperPRL1961,DeGennesRMP1964}.
At low temperature, the coherence length of NbN is $\xi(0)\approx 4.5\,\mathrm{nm}$ \cite{cockalingam08}, but close to the transition (let's say, for $(T_c-T) < 100$ mK)  $\xi(T)=\xi(0)/[1-(T/T_{c})^4] \geq 50 \, \mathrm{nm}$, indicating that our films are firmly in the Cooper limit. Since NbN is well described by the BCS theory in the strong coupling regime, we can use the strong-coupling theory for proximity effect in the Cooper limit \cite{Silvert75}. As a first approximation we can assume that both the characteristic temperature $\Theta$ (representative of the phonon spectrum and thus related to the Debye temperature) and the Coulomb pseudopotential $\mu^*$ are unaffected by the applied electric field. In this case they can thus be obtained from literature \cite{cockalingam08} and the model of Ref. \onlinecite{Silvert75} gives for the critical temperature of the compound slab:}

{\color{black}\vspace{-5mm}
\begin{equation}
T_{c,comp} = \frac{\Theta}{1.45}\exp\left[-\frac{1+\left\langle \lambda\right\rangle}{\left\langle \lambda\right\rangle -\mu^{*}}\right]
\label{eq:Tccomb}
\end{equation}
where
\begin{equation}
\langle\lambda\rangle = \frac{\lambda_{s}N_{s}d_{s}+\lambda_{b}N_{b}d_{b}}{N_{s}d_{s}+N_{b}d_{b}}= \beta_{s}\lambda_{s}+\beta_{b}\lambda_{b}.
\end{equation}
}%
Here, the subscripts $s$ and $b$ refer to surface and bulk, $N_{s,b}$ are the densities of states at the Fermi level and $d_{s,b}$ the thicknesses of the layers, such that $d_{s}+d_{b}=t$. As for the electron-phonon coupling strength $\lambda_{s}$, the simplest modification due to the induced charge density is $\lambda_{s}=\lambda_{b}\cdot N_{s}/N_{b}$, $\lambda_{b}$ and $N_b$ being calculated from the unperturbed $T_c$ through the McMillan equation, and via density functional theory (DFT), respectively. The only remaining unknown quantity is the surface DOS at the Fermi level $N_{s}$.
In general, the shift of the Fermi level is determined by the volume density of induced carriers $\Delta n_{3D}$, while {\color{black}Double-Step Chronocoulometry} is able to measure the surface charge density $\Delta n_{2D}=\int_{0}^{t}\Delta n_{3D}(z)dz$ \cite{daghero12}. An ansatz about how the volume charge density distributes across the thickness is thus required to determine $N_s$.

\begin{figure}
\includegraphics[angle=-90,width=0.9\columnwidth]{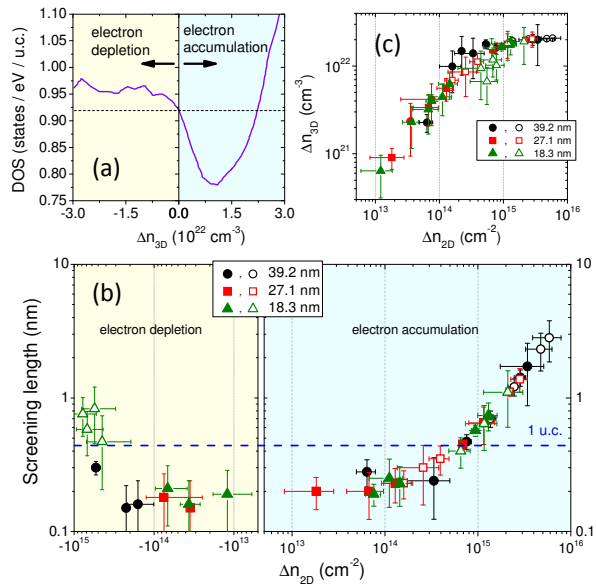}
\caption{(a) Density of states (DOS) of NbN as a function of the volume density of induced electrons $\Delta n_{3D}$ (i.e. $\Delta n_{3D}=0$ corresponds to native NbN, without gating). (b) Thickness of the perturbed surface layer $d_s$ (that can be roughly assumed to be the screening length $\xi_s$) vs. $\Delta n_{2D}$ for both electron accumulation and depletion. The horizontal dashed line indicates the size of one unit cell of NbN. (c) Absolute value of the volume density of induced electrons (in the surface layer) $\Delta n_{3D}$ as a function of $\Delta n_{2D}$.}\vspace{-5mm}
\label{figure:4}
\end{figure}

Since within the model in Ref. \cite{Silvert75} the two layers of the compound slab are homogeneous, we choose a step profile for $\Delta n_{3D} (z)$, i.e. we assume the induced charge to be uniformly distributed in a thickness $d_{s}$, which is an adjustable parameter of the model. Clearly, $d_s$ is connected to the screening length $\xi_s$ even though it refers to a simplified step-like $z$-profile distribution of volume charges. For any given value of $\Delta n_{2D}$, the choice of $d_s$ determines $\Delta n_{3D}$ and consequently: i) (by DFT calculations) the shift of the Fermi level and the perturbed DOS at the surface, $N_s$ (see Fig. \ref{figure:4}a); ii) the electron-phonon coupling strength $\lambda_s$; iii) the value of $T_{c, s}$, and, finally, the critical temperature of the compound slab $T_{c,comp}$ which has to agree with the experimental $T_c$.

The values of $d_s$ (that we can take as an \emph{effective} electrostatic screening length) that allow fitting the experimental $T_c$ shifts, determined through an iterative procedure, are plotted as a function of $\Delta n_{2D}$ in Fig. \ref{figure:4}b. Symbols of different shape refer to different film thicknesses $t$ \footnote{We excluded from our analysis the data for $t=9.5$ nm as we deem the measured  $T_c$ shift not to be reliable enough due to the strong hysteresis of the field effect (see Supplemental Material for details).}.
It becomes immediately clear that the \emph{effective} screening length does not depend on $t$, which is quite reasonable, but must vary with $\Delta n_{2D}$. Let us focus on the electron accumulation side, where the trend is clearer. In the low-carrier density region, $d_s$ roughly agrees with the Thomas-Fermi screening model if the density of quasiparticles present at $T \simeq T_c$ is used; but already at $7\times10^{14}\, \mathrm{cm^{-2}}$ it becomes as large as one unit cell ($4.4$ {\AA}). Without this increase in $d_s$, the volume charge density $\Delta n_{3D}$ would become so large that the Fermi level would be shifted well beyond the local minimum in the DOS (see fig. \ref{figure:4}a), resulting in an \emph{increase} in the DOS $N_s$ and thus in a \emph{positive} $\Delta T_c$, which is not the experimental result. For larger values of $\Delta n_{2D}$, $d_s$ further expands, finally reaching 4-5 unit cells. For $\Delta n_{2D}>5 \times 10^{14} \mathrm{cm^{-2}}$ the dependence of $d_s$ on $\Delta n_{2D}$ is remarkably linear. Note that the increase in $d_s$ is not fast enough to keep the volume density of induced electrons $\Delta n_{3D}$ constant; indeed, this quantity increases as well, as shown in Figure \ref{figure:4}c, and tends to saturate around $2 \times 10^{22} \, \mathrm{cm^{-3}}$.

These results indicate that the volume density of induced charges cannot exceed $2 \times 10^{22} \, \mathrm{cm^{-3}}$, and that the thickness of the surface layer departs from a Thomas-Fermi value (see fig. \ref{figure:4}b) when $\Delta n_{3D}$ approaches this limit (see fig.\ref{figure:4}c) as if the surface layer of thickness $\approx \lambda_{TF}$ was unable to accommodate all the induced charges. To look for an explanation of this effect, one certainly has to abandon the Thomas-Fermi approximation: In this high charge-density regime the assumptions of weak perturbation and linear response are no longer valid since the surface potential $\phi(z=0)$ does no longer fulfill the condition $|e\phi(z=0)| \ll E_{F}$. The screening theory beyond the linear regime \cite{Chazalviel} correctly explains the observed increase of the screening length up to about 3.6 {\AA} when $\Delta n_{2D} \simeq 5 \times 10^{14} \, \mathrm{cm^{-2}}$ (see Supplemental Material for details), but above this doping value the appropriate theory is lacking.


In summary, we have experimentally proven that a \emph{surface} charge induction by electrochemical gating can give rise to modifications of the \emph{bulk} superconducting properties (and not only of the surface ones). This is true, surprisingly, in conventional BCS-like superconductors with a large electronic screening, and can be explained in terms of proximity effect between the surface layer and the underlying part of the sample. We have also unveiled an increase in the effective electronic screening length, that departs from the Thomas-Fermi value and increases, suggesting the existence of an upper limit for the volume charge density.
These findings severely impact the study of the effects of EDL gating on high carrier density systems in general, and metallic superconductors in particular.

\vspace{3cm}
\begin{center}
  \Large{Supplemental material}
\end{center}

\normalsize

\section{Composition of the PES}
We employed two different PES compositions in different sets of experiments, in order to rule out any influence of the specific electrolyte formulation on the measured effects: i) a Li-TFSI based PES we have already used in the experiments on noble metals \cite{daghero12,tortello13}; ii) a similar PES where the Li-TFSI-PEGMA mixture was substituted by pure 1-Butyl-1-methylpiperidinium bis(trifluoromethylsulfonyl)imide ionic liquid.

\section{Ion milling}
All the measurements described in the present paper were made on two different NbN devices, whose thickness was progressively reduced by Ar-ion milling. In other words, once the FE measurements for a given thickness were completed, we removed the PES by rinsing the device in ethanol, then carefully cleaned the surface via sonication in acetone and ethanol. The physical removal of  NbN  was then performed by an Electron Cyclotron Resonance Plasma System operating at 2.45 GHz, in which the sample was placed 10 cm above the plasma source. The milling was carried out at an Ar pressure of $1.0 \times 10^{-3}$ mbar and, using an extraction voltage of 400 V and an anode current density of 1.5 mA cm$^{-2}$, an etching rate of 0.14  nm s$^{-1}$ was obtained. The whole sample area was exposed to the ion flux; since the ions etch the MgO substrate more efficiently than the NbN film, the milling process was calibrated on an unpatterned film with the help of a profilometer (by measuring the height difference between the exposed area and a region protected by a polymeric mask). The exposure time was optimized to allow for a thickness reduction of about 10 nm in each milling step. After each milling step, we characterized the surface morphology of the device via AFM. This was to ensure that the quality of the surface did not get degraded in a significant way.\newline

\begin{figure}[ht]
\includegraphics[width=0.90\columnwidth]{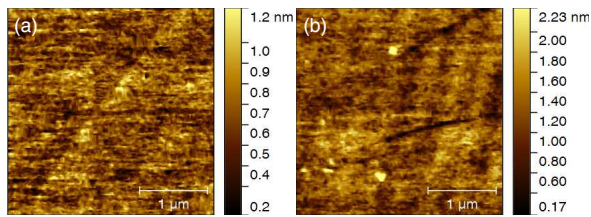}
\caption{AFM topography maps of the active channel of the same NbN device. Panel (a) shows the surface of the pristine film. Panel (b) shows the surface of the film after three ion milling steps. Both maps are 3 $\mu$m $\times$ 3 $\mu$m in lateral dimension. Scale bars on the left show the AFM heigth signal for the two maps.}\label{figure:AFM}
\end{figure}

Figure \ref{figure:AFM} shows two 3 $\mu$m $\times$ 3 $\mu$m topography maps of the same device. The map on the left was measured on the pristine device, before any ion milling steps ($t = 39.2 \pm 0.8$ nm). The map on the right was measured after three ion milling steps, i.e. for the smallest film thickness reached in our experiments ($t = 9.5 \pm 1.8$ nm). The pristine sample appears to be almost atomically flat, with a terrace clearly visible across the map. The milled sample presents a somewhat rougher surface, as is to be expected. Both maps feature a surface roughness below $1$ nm. In order to obtain a reliable estimation of the roughness of our devices, we measured several topography maps in different regions of the devices and calculated the arithmetic average of the 2D roughness profile ($R_a$) for each map. We then averaged the $R_a$ themselves for each milling step. By this method, the pristine devices showed $R_a = 0.32 \pm 0.06$ nm. After the first ion milling step, the surface roughness increased to $R_a = 0.45 \pm 0.15$ nm. The second and third ion milling steps did not degrade the surface further.

\section{Details of transport measurements}
\begin{figure}[ht]
\includegraphics[width=0.80\columnwidth]{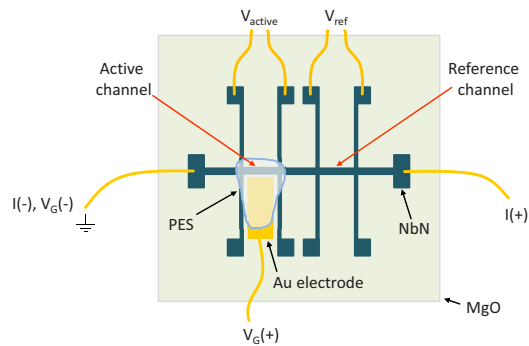}
\caption{Sketch of a gated NbN device. The edges of the PES, that define the area of the device capacitively coupled to the gate electrode, are highlighted in blue for clarity.}\label{figure:Figure_SUPPL_device}
\end{figure}

Transport measurements were performed in a Cryomech\textregistered{} ST-403 3K pulse-tube cryocooler. The source-drain current $I_{SD}$ was provided by a Keithley 6221 source and was kept sufficiently small (a few tens of $\mu$A) so as to avoid heating in the film (even in the thinnest ones). The possible thermoelectric effects were eliminated by inverting the current within each measurement \cite{daghero12}. The longitudinal voltage drops along the active and reference channels, $V^{active}$ and $V^{ref}$, were measured by a low-noise Keithley 2182A nanovoltmeter, and the sheet resistance $R_{\Box}$ of each channel was calculated accordingly.
The measurements were performed during both cool-down to 2.7 K and subsequent warm-up to room temperature. However, since the mobility of the ions in the PES is strongly damped below the glassy transition of the PES, every change in $V_G$ had to be performed at high temperature (above 230 K) and the only way to detect a $T_c$ shift was to compare resistance curves obtained in different thermal cycles. To maximize the reproducibility of the $T_c$ measurement, we only considered the heating curves for the analysis, since the warming up was quasistatic. Nevertheless, the variation in $T_c$ measured in the same channel in different heating processes was comparable to the shift due to EDL gating. It is precisely to overcome this problem that we measured at the same time the resistance of the active (gated) and of the reference (ungated) channels. {\color{black}Figure \ref{figure:Figure_SUPPL_device} presents a sketch of the full device, showing both the region of the film that is capacitively coupled to the gate electrode (active channel) and the region that is not (reference channel).
A further possible source of systematic error in the measurements lies in the fact that, in general, the modulation to the carrier density induced by EDL gating may be slightly temperature-dependent \cite{Ye10,li_nature16}. Thus, the measurement of the induced charge performed at room temperature by Double-Step Chronocoulometry may not correctly estimate the carrier density modulation affecting the superconducting properties at low temperature. While this is a known issue in EDL-gated 2D materials and may be due to disorder \cite{Ye10} or charge-ordering \cite{li_nature16}, we don't expect it to play a significant role in metallic thin films. Indeed, if the density of extra charge carriers introduced by the gate was temperature-dependent, we would expect the resistance modulation in the normal state to be temperature-dependent as well. Instead, our earlier results on gold \cite{daghero12} and other noble metals \cite{tortello13} showed that the $R$ vs. $T$ curves are rigidly shifted by the applied gate voltage, with no sign of a further temperature-dependent component. We can thus discard the possibility of a significant temperature dependence of the induced carrier density in our metallic films.}

\section{Reversibility of the $T_c$ shift}
\begin{figure}[ht]
\includegraphics[height=0.8\columnwidth,angle=-90]{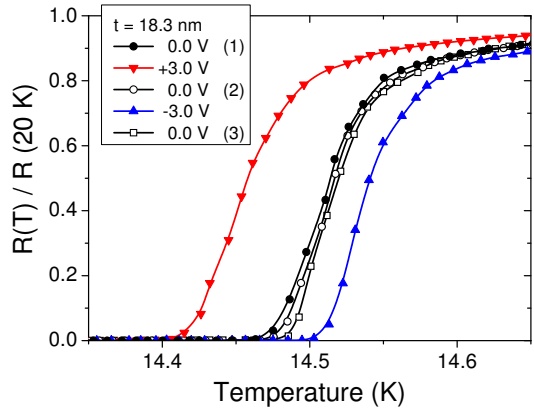}
\caption{Five $R_{\Box}(T)$ curves measured subsequently in the  18.3 nm thick device. The superconducting transition temperature at $V_G=0$ returns to the original value after applying and removing a gate voltage $V_G = +3$ V and $V_G = -3$ V, confirming the complete reversibility of the effect and thus its purely electrostatic origin.}\label{figure:rev}
\end{figure}

In order to check the purely electrostatic nature of the $T_c$ modulation, i.e. to ensure that the $T_c$ shift was completely reversible upon removal of the gate voltage, we performed measurements of $R_{\Box}(T)$ at zero gate voltage before and after the application of each $V_G$. This was necessary due to the ubiquitous possibility that electrochemical reactions occur at the electrolyte/electrode interface, leading to permanent or semipermanent modifications of the material under study \cite{Piatti16}. Figure \ref{figure:rev} shows five $R_{\Box}(T)$ curves measured subsequently in the 18.3 nm thick device, with $V_G=$ 0, +3 V, 0, -3 V, 0. Note that the horizontal scale is the absolute temperature (i.e. not the referenced one). The three curves recorded at $V_G =0$ fall exactly on top of each other, the very small shift being only due to the different heating runs and thus being completely corrected if the referenced temperature is used. {\color{black}The longest consecutive series of measurements where the reproducibility of the $V_G = 0$ curve was checked featured seven different $V_G$ values, for a total of 15 thermal cycles.} This reproducibility confirms the absence of significant electrochemical interaction between the PES and the NbN thin film even when a gate voltage of $\pm 3$ V is applied.

The 9.5 nm thick device was anomalous in this sense. In that case, the superconducting transition temperature did not go back to the original value upon removal of the gate voltage.  Instead, we observed a positive irreversible $T_c$ shift in addition to the usual (reversible) electrostatic effect, independently of the sign of the gate voltage. Remarkably, the transition profiles remained unaffected. By using the technique described above and shown in Fig.\ref{figure:rev}, we were able to separate the irreversible effect from the reversible one. The $T_c$ shifts shown in Fig. 2 and 3 of the main text are only the \emph{reversible} ones; conservatively, we only reported the data for which the reversible effect was much greater than the irreversible one. {\color{black}While a quantitative explanation for this slow relaxation in the $T_c$ of NbN at small film thicknesses is beyond the scope of this paper, we can provide a qualitative assessment of the problem. It is very well known that the NbN and MgO unit cells feature a mismatch of $\sim 3$\% [S1]. This can result in a native strain in NbN films grown on MgO for film thicknesses below $\sim 50$ nm [S1]. This mismatch can also lead to the formation of slightly misaligned three-dimensional islands during the growth process, with structural defects such as dislocations, vacancies and voids in films up to $\sim 15$ nm thick [S1]. By ion-milling our films, we eventually expect to directly expose this damaged and strained region to the PES and the intense electric field built up in the EDL. We thus interpret the progressive shift in $T_c$ upon several thermal cycles under the presence of the electric field as a parasitic electrostriction effect that induces a progressive relaxation of the strained NbN in close proximity to the mismatched MgO substrate. This effect is thus completely negligible for films of larger thickness, that would be significantly less defective and feature a lower strain, as also demonstrated by the reversibility of the $T_c$ shifts that we indeed measured in these thicker films.} For the same reason, we decided not to apply the model for the proximity effect to the values of $\Delta T_c$ measured in the 9.5-nm thick device (see Figure 4 of the main text).
Since the irreversible $T_c$ shift was likely to become worse at even smaller thicknesses, we opted not to thin down our devices further. Another reason why we did not explore thicknesses below 9.5 nm is that we were not confident that the film would retain a sufficient electrical continuity and a proper surface quality upon further ion-milling steps.

\section{Density functional theory calculations of the $\bold{\mathrm{NbN}}$ density of states}
The total density of states (DOS) of {\color{black}the bulk crystalline} NbN has been ab-initio calculated by density functional theory (DFT) employing the all-electron full-potential linearised augmented-plane-wave (FP-LAPW) method as implemented in the Elk code [S2]. In order to describe the exchange and correlation energy term we adopted the Generalised Gradient Approximation (GGA) as developed in Ref. [S3]. The Brillouin zone was sampled with a 24 $\times$ 24 $\times$ 24 mesh of $k$-points and the convergence of self-consistent field calculations was attained with a Kohn-Sham effective potential tolerance of 1$\times 10^{-9}$ and a total energy tolerance of 1$\times 10^{-8}$ Hartree. The DOS close to the Fermi energy ($E_F$) has been calculated with an energy resolution of 3 meV between -0.6 and +0.6 eV from $E_F$. The electronic band structure in the $k$-space and the total DOS (calculated in the range from -13.6 eV to 13.6 eV with respect to $E_F$ and with an energy resolution of 5 meV) are in very good agreement with the results in the literature [S4,S5].

\section{Further details about the proximity-effect model}
As explained in the text, we used a strong-coupling model for the proximity effect \cite{Silvert75} to calculate the relationship between the experimental critical temperature of the film, treated as a compound slab, and the surface density of induced charges $\Delta n_{2D}$. To do so, we assumed the induced carrier density per unit volume in the surface layer (i.e. $\Delta n_{3D}(z)$) to follow a step-like profile as a function of the depth $z$. We called $d_{s}$ the thickness of the charge-induction layer.
According to that model, for a given value of $\Delta n_{2D}$ and $d_{s}$, the computed $T_{c,comp}$ of the compound slab varies with the total film thickness $t$ in a way that qualitatively agrees with the experimental trend at low and moderate induced carrier densities: The difference between $T_{c,comp}$ and the unperturbed critical temperature is strongly reduced by increasing $t$, eventually becoming comparable with the experimental uncertainty for $t>100$ nm.

\begin{figure}[ht]
\vspace{0cm}
\includegraphics[width=0.9\columnwidth,angle=-90]{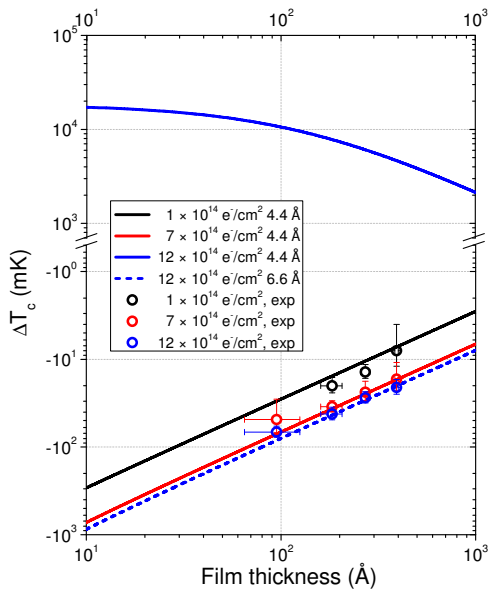}
\caption{Critical temperature shift, $\Delta T_c$, vs. film thickness for three selected
values of accumulated electron density ($\Delta n_{2D} = 1,\,7,\,12\times 10^{14} \, \mathrm{cm^{-2}}$).
Solid (dashed) lines result from the proximity effect model for a
surface layer thickness of 4.4 {\AA} (6.6 {\AA}).}\label{figure:Tcvsthickness}
\end{figure}

In Fig. \ref{figure:Tcvsthickness} the experimental values of the $T_c$ shift are reported as a function of the film thickness $t$ for three different values of $\Delta n_{2D}$, i.e. $1\times 10^{14}\,\mathrm{cm^{-2}}$ (black symbols), $7 \times 10^{14}\,\mathrm{cm^{-2}}$ (red symbols) and $12\times 10^{14}\,\mathrm{cm^{-2}}$ (blue symbols). In the first two cases (black and red solid lines), the experimental data can be reproduced fairly well by using the same value of $d_{s}=4.4$ {\AA} [S6]  -- which, by the way, corresponds to the height of the unit cell.  In this range of $\Delta n_{2D}$, the induced volume charge density $\Delta n_{3D}$ in the surface layer results in a displacement of the Fermi level that corresponds to a reduction in the DOS (see Fig. 4a in the main text). At the  highest value of $\Delta n_{2D}$, however, the same $d_{s}$ would lead to such a high value of $\Delta n_{3D}$ that the Fermi level would be pushed beyond the local minimum, leading to an \emph{increase} in the DOS, and a consequent \emph{increase} in the $T_c$, contrary to the experimental finding. The blue solid line in Fig.\ref{figure:Tcvsthickness} represents the $\Delta T_c$ vs. $t$ calculated in this case.

To obtain an agreement with the experimental data for $\Delta n_{2D} = 12\times 10^{14}\,\mathrm{cm^{-2}}$, the value of $d_{s}$ must be increased to 6.6 {\AA} (dashed blue line).
It is also worthwhile to note that, according to the model, if the change in the critical temperature of the surface layer, $T_{c, s}$, was larger than a few Kelvin, the resulting $T_{c, comp}$ would be comparable with  $T_{c, s}$; in other words, for large modifications of $T_{c, s}$ the proximity effect would play almost no role in determining the final $T_{c, comp}$. This result is in agreement with recent findings on EDL-gated FeSe thin flakes \cite{shiogai15,lei16}.

\section{Electrostatic screening beyond the linear regime for a planar distribution of charges in a metal}

In order to try to explain the anomalous screening length determined at very high induced charge density by the proximity-effect model we applied the screening theory beyond the linear regime \cite{Chazalviel}. In this case the standard approximation at the basis of the linear regime in a metal, i.e. $|e\phi| \ll E_F$ (where $\phi$ is the electrostatic potential of the induced charge and $E_F$ is the Fermi energy of the metal), doesn't hold any more. For a planar distribution of charges in a degenerate Fermi gas this theory predicts a general equation that involves the integral of the density of states $N(E)$ of the metal (see Ref. \onlinecite{Chazalviel}, pag. 46, eq. II-7):

\begin{equation}
\pm\sqrt{\frac{2}{\epsilon_{r}\epsilon_{0}}}z = \int_{\phi_{S}}^{\phi}\frac{d\varphi}{\left(\int_{0}^{\varphi}-ed\varphi'\int_{E_{F}+e\varphi'}^{E_{F}}N(E)dE\right)^{1/2}}
\label{eq:intN(E)}
\end{equation}
where $\epsilon_r$ is the relative permittivity, $z$ is the coordinate perpendicular to the planar charge distribution, directed toward its inner part, and $\phi_S$ is the potential at its surface, i.e. $\phi_S = \phi(z=0)$. We know the behaviour of $N(E)$ close to $E_F$ from ab-initio DFT calculations. Thus by inserting in eq. \ref{eq:intN(E)} a linear approximation for $N(E)$ in the electron accumulation region (see fig. 2a of Ref. \onlinecite{Piatti16}), $N(E) = N(E_F)-k(E-E_F)$, we can integrate equation \ref{eq:intN(E)} obtaining:
\begin{equation}
\phi(z\geq0)=\frac{a}{b}\left\{ 1-\tanh^{2}\left[\frac{\sqrt{a}}{2}\sqrt{\frac{2}{\epsilon_{r}\epsilon_{0}}}z+\tanh^{-1}\left(\frac{\sqrt{a-b\phi_{S}}}{\sqrt{a}}\right)\right]\right\}.
\label{eq:phi(x)}
\end{equation}

Here $a = e^2 N(E_F) / 2$ and $b = e^3 k / 6$ are known from DFT calculations (see also [S7] for comparison). As a consequence the $z$ dependence of the potential $\phi(z)$ is fully determined by the choice of the surface potential $\phi_{S}$. This choice is not arbitrary since we can select the $\phi_{S}$ value in order to match the experimental induced surface charge density $\Delta n_{2D}$ that is given by:

\begin{equation}
\Delta n_{2D}=\int_{0}^{\infty}\frac{\epsilon_{r}\epsilon_{0}}{e}\frac{d^{2}\phi(z)}{dz^{2}}dz.
\label{eq:Deltan}
\end{equation}

In the framework of this theory beyond the linear regime we calculate the screening length as the $z$ coordinate at which the volume charge density $\Delta n_{3D}=\left(\epsilon_{r}\epsilon_{0}/e\right)\left(d^{2}\phi\left(z\right)/dz^{2}\right)$ is $1/e$ of its surface value $\Delta n_{3D}(z=0)$. When $\Delta n_{2D} \lesssim 1 \times 10^{14} \mathrm{cm^{-2}}$ this length coincides (within 10 \%) with the screening length given by the standard Thomas-Fermi theory $\lambda_{TF}=\sqrt{\epsilon_{r}\epsilon_{0}/2a} \simeq 0.2$ nm and nicely corresponds to the constant value determined from the experiments (see Fig. 4b of the main text). At the increase of $\Delta n_{2D}$ the screening length in the non-linear regime progressively increases reaching 0.36 nm when $\Delta n_{2D} = 5.35 \times 10^{14} \mathrm{cm^{-2}}$, again in very good agreement with the experimental results shown in Fig. 4b. For larger values of $\Delta n_{2D}$ this theory is no more able to explain the observed linear increase of the screening length that thus remains an open problem.

\vspace{5mm}\small

\noindent
[S1] L. Hultman, L. R. Wallenberg, M. Shinn, and S. A. Barnett, \textit{J. Vac. Sci. Technol. A} \textbf{10}, 1618 (1992)

\noindent
[S2] Dewhurst, K. et al. The Elk FP-LAPW Code. http://elk.sourceforge.net/

\noindent
[S3] Yingkai Zhang and Weitao Yang, \emph{Phys. Rev. Lett.} \textbf{80}, 890 (1998)

\noindent
[S4] AFLOW Automatic-FLOW for material discovery, Center for Material Genomics, Material Science, Duke University, 2016. http://www.aflowlib.org/

\noindent
[S5] T. Amriou et al., \emph{Physica B} \textbf{325} 4656 (2003)

\noindent
[S6] This value of $d_{s}$ is just a guess. For $\Delta n_{2D}=1 \times 10^{14} \,\mathrm{cm^{-2}}$ it differs from that reported in Figure 4 of the main text, which was instead determined by the iterative best-fit procedure described there.

\noindent
[S7] S. Blackburn, M. C\^{o}t\'{e}, S. G. Louie and M. L. Cohen, \textit{Phys. Rev. B} \textbf{84}, 104506 (2011)

\end{document}